\documentstyle[11pt,epsf]{article}

\begin{document}

\begin{flushright}
SU-4240-632 \\
May 1996
\end{flushright}

\begin{center}
\vspace{24pt}

{\Large \bf Minimal Dynamical Triangulations of Random Surfaces}

\vspace{24pt}

{\large \sl Mark J.~Bowick, Simon M.~Catterall  \\
 {\rm and}  Gudmar Thorleifsson} \\ 
\vspace{6pt}
Department of Physics, Syracuse University \\
Syracuse, NY 13244-1130, USA

\vspace{15pt}

\begin{abstract}
We introduce and investigate numerically a {\em minimal} class of
dynamical triangulations of two-dimensional gravity on the sphere in
which only vertices of order five, six or seven are permitted. We show
firstly that this restriction of the local coordination number, or
equivalently intrinsic scalar curvature, leaves intact the fractal
structure characteristic of generic dynamically triangulated random
surfaces.  Furthermore the Ising model coupled to {\em minimal}
two-dimensional gravity still possesses a {\em continuous} phase
transition.  The critical exponents of this transition correspond to
the usual KPZ exponents associated with coupling a central charge
$c=\frac{1}{2}$ model to two-dimensional gravity.
\end{abstract}

\end{center}

\vspace{20pt}

That two-dimensional ($2D$) quantum gravity can be regularized and
studied using {\it dynamical triangulations} (DT) is well known
\cite{generic} and has lead to extensive study of the properties of
dynamically triangulated random surfaces (DTRS).  Indeed, many of the
properties of critical spin systems on such lattices have been shown
to follow from continuum treatments of central charge $c<1$ theories
coupled to $2D$-quantum gravity.  Furthermore, this formulation has 
the
advantage that it renders tractable the calculation of many features
of these theories which are inaccessible to continuum
methods. Foremost amongst these are questions related to the quantum
geometry and fractal structure of the typical $2D$ manifolds appearing
in the partition function \cite{frac}.

It has often been speculated that some of these models should also
find a realization in condensed matter physics as models of membranes
and/or interfaces with {\it fluid} in-plane degrees of freedom.  The
problem in trying to map such problems onto dynamically triangulated
models, however, has always been the occurrence of large vertex
coordination numbers in an arbitrary random graph {--} real condensed
matter systems typically place a cut-off on the number of interactions
allowed by the microscopic degrees of freedom.  It is possible that
such cut-offs are unimportant in the critical region of these
theories. As a crude analogy we note that the short-distance
differences between different discrete polygonal decompositions of
two-dimensional gravity are, without fine-tuning, irrelevant in the continuum 
limit \cite{generic}.

In this letter we present numerical results for a new model (called
the minimal dynamical triangulation or MDT model) which is based on
random triangulations but with an additional constraint: the
coordination number of any vertex in the lattice can only be five, six
or seven. Such minimal models implement a severe form of the
microscopic interaction cut-off required by condensed matter systems
and correspond in gravitational language to the inclusion of only
three possible states for the local curvature ${-2\pi/7,\,
0,\,2\pi/5}$.  (The scalar curvature at a vertex with coordination
number $q_i$ is given by $R = 2\pi \frac{6-q_i}{q_i}$.) We have
studied two systems: pure gravity as modelled by pure (minimal)
triangulations and a set of Ising spins coupled to a dynamical lattice
generated by minimal triangulations.  We show in both cases that the
critical behavior of the system appears to be identical to the
conventional DTRS models {--} yet another striking indication of
universality.

The dynamically triangulated version of $2D${--}gravity 
is given by the partition function
\begin{eqnarray}
\label{e21}
Z(\mu) &= &\sum_{A} {\rm e}^{\textstyle -\mu A} Z(A), \\ 
Z(A)  &=  &\sum_{T \in {\cal T}} Z_M \; .
\end{eqnarray}
$Z(A)$ is the fixed area partition function, $\mu$ the
cosmological constant and $\cal T$ the class of triangulations 
summed over. $Z_M$ represents the
partition function of some generic matter field living on the
surface.  A given choice of $\cal T$ corresponds to a particular
discretization of the surfaces.  Two classes 
commonly used are combinatorial and degenerate triangulations,
${\cal T}_C$ and ${\cal T}_D$.  In the former triangles are 
glued together along edges so as to form closed manifolds, with 
the constraints that any two triangles can only have one edge
in common and no triangle can be glued onto itself.  For 
degenerate triangulations these constraints are relaxed {--} this
amounts to allowing tadpoles and self-energy diagrams in the dual
representation to triangulations.  Clearly ${\cal T}_C
\subset {\cal T}_D$.

A priori different classes ${\cal T}$ do not have to lead to the same
theory, although the two aforementioned do, at least
for the case of Ising matter 
where the models are exactly soluble \cite{solv}.  
Clearly this would no longer 
be
true if ${\cal T}$ only contained {\it one} triangulation {--} an
extreme limit.  The class of triangulation ${\cal T}_M$ we investigate
in this letter is a subclass of ${\cal T}_C$ defined by allowing only
vertices of order 5, 6 or 7.  This is as far one can go in suppressing
curvature fluctuations on the surface while still retaining its
dynamical (fluid) nature.

We start by investigating the minimal model in the absence of matter
($Z_M = 1$ in Eq.\ 2).  To see if this model belongs to the same
universality class as pure gravity with triangulation space ${\cal
T}_C$ or ${\cal T}_D$ we study its critical behavior numerically.
More precisely we measure the string susceptibility exponent
$\gamma_s$, defined through the scaling $Z(\mu) \sim (\mu_c -
\mu)^{2-\gamma_s}$.  The method we employ is that of calculating the
distribution of {\it minimal baby universes} $n(B)$ on the ensemble of
triangulations \cite{baby2}.  A baby universe of area $B$ is a region
of the triangulation joined to the bulk solely by a minimal neck.  For
the class ${\cal T}_M$ (and ${\cal T}_C$) a minimal neck is a loop
consisting of three links.  Using the expected asymptotic behavior of
the fixed area partition function, $\; Z(A) \sim \exp (\mu_c A)
A^{\gamma_s - 3}$, a simple argument relates $n(B)$ to $\gamma_s$
\cite{baby2,foot1}; 
\begin{equation}
\label{e22}
n_A(B) \sim  \left [ B(A-B) \right ]^{\gamma_s - 2},
\;\;\;\; A \rightarrow \infty.
\end{equation} 

We have simulated the fixed area partition function $Z(A)$, using
Monte Carlo methods, for triangulations consisting of 1000 and 4000
vertices \cite{foot2}.
Henceforth we will use the number of
vertices as our measure of the area $A$. In Fig.\ 1 we show the
measured distribution $n(B)$ for $A = 1000$.  For comparison we
include the corresponding distribution measured on the class of
combinatorial manifolds ${\cal T}_C$. Fig.~1 also includes a third
distribution $(c)$ whose significance will be addressed shortly.  The
first thing we notice is that these distributions are very different
for $B < 100$ {--} in particular there are no small baby universes for
the class ${\cal T}_M$ as the curvature restriction puts a lower limit
on the size of baby universes; our construction effectively smoothens
the triangulation locally.

\begin{figure}
\epsfxsize=4.5in  \centerline{ \epsfbox{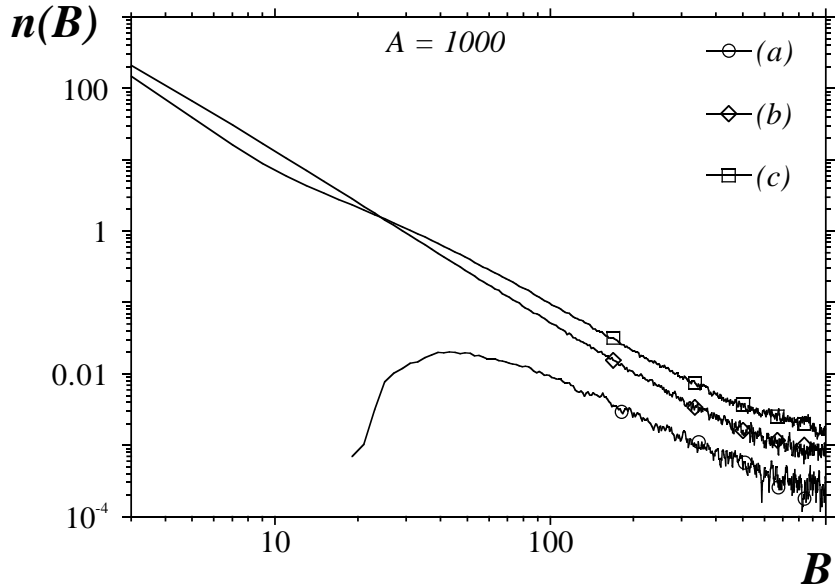}}
\caption{The measured distributions of baby universes for
 {\it (a)} the minimal class of triangulations, {\it (b)}
 combinatorial triangulations and {\it (c)} triangulations
 obtained by node decimation from {\it (a)}.  All measurements
 are for triangulations of 1000 nodes.}
\label{fig:bu1}
\end{figure}

For larger baby universes the distributions look much more  
alike. To measure $\gamma_s$ we fit the tail of $n(B)$ to
Eq.~3. This gives $\gamma_s = -0.64(5)$ ($A = 1000$)
and $-0.53(3)$ ($A = 4000$). This agrees with 
the exact value for pure gravity $\gamma_s = -0.5$.
The corresponding value obtained using ${\cal T}_C$ is
$\gamma_s = -0.501(4)$ ($A = 1000$) and $-0.504(4)$ ($A = 4000$).
It is easy to understand why the values for minimal triangulations
are less accurate {--} finite-size effects are
more pronounced since more data for small values of $B$
has to be excluded before we reach the asymptotic regime.
Yet the result indicates that the minimal triangulated model is 
in the same universality class as the full-blown DTRS model.

Additional evidence for this claim is obtained by applying a recently
proposed Monte Carlo renormalization group method for blocking
dynamical triangulations ({\it node decimation} \cite{ndec}) to the
model.  Each triangulation is blocked by removing vertices at random
and in the process the restrictions on the curvature are dropped so
that the model flows into the wider class of combinatorial
triangulations.  In this way we can ascertain whether the model after 
blocking is
the same as that obtained with the triangulation class ${\cal T}_C$
{--} in particular we would like to know if the RG transformation
generates a flow towards a fixed point, accessible from combinatorial 
triangulations, and corresponding to pure gravity.

Curve $(c)$ of Fig.~1 is the measured distribution $n(B)$ for surfaces
of 1000 vertices obtained from surfaces of 4000 vertices by node
decimation.  It is apparent that this distribution is much closer to
the distribution for combinatorial triangulations than the minimal
one.  This we take as an indication that the model flows towards the
same non-trivial pure gravity fixed point as combinatorial manifolds.
This evidence can be strengthened by measuring the exponent $\gamma_s$
on the set of triangulations obtained with blocking.  This we show in
Table~\ref{tab1} for various degrees of blocking.  Agreement with the
pure gravity value clearly improves with blocking.  This confirms that
the model, after blocking, is closer to the pure gravity fixed point.

\begin{table}
\begin{center}
\caption{Measured values of $\gamma_s$ for minimal
dynamical triangulations (without matter) after
applying varying levels of node decimation with
a blocking factor of $b = 2$.}
\vspace{0.1in}
\begin{tabular}{ccccc} \hline
 & \multicolumn{2}{c}{$A^{(0)} = 1000$} 
 & \multicolumn{2}{c}{$A^{(0)} = 4000$} \\  \cline{2-3} \cline{4-5}
$A^{(k)}$ &  $k$  &  $\gamma_s$  &  $k$  &  $\gamma_s$   \\ \hline
4000    &       &              &  0    &  -0.530(26)   \\
2000    &       &              &  1    &  -0.544(32)   \\
1000    &  0    &  -0.644(48)  &  2    &  -0.530(18)   \\
500     &  1    &  -0.619(26)  &  3    &  -0.504(9)    \\
250     &  2    &  -0.574(49)  &  4    &  -0.478(36)   \\ \hline
\end{tabular}
\label{tab1}
\end{center}
\end{table}

The RG flows above can be visualized by looking at the expectation
values of a subset of operators such as 
\begin{equation}
 O_1^{(k)} = \left < \sum_i
(q_i^{(k)}-6)^2 \right > \;\;\;\; {\rm and} \;\;\;\;
 O_2^{(k)} = \left < \sum_{<ij>}
(q_i^{(k)}-6) (q_j^{(k)}-6) \right >, 
\end{equation}
where $q_i$ is the curvature of
a node $i$, $k$ is the blocking level and the first sum is over nodes
while the second is over nearest neighbors.  Different trajectories in
the $(O_1,O_2)${--}plane can be accessed by adding an irrelevant
operator to the bare action and varying its coupling constant.  We
have chosen $\; \alpha \sum_i \log q_i \;$.  This is shown in Fig.~2.
By changing the coupling constant $\alpha$ we effectively change the
trajectories identifying the location of the non-trivial fixed point.
And indeed the trajectory for the minimal triangulations flows
directly towards the pure gravity fixed point before it veers off and
follows a universal trajectory into a zero-volume sink.

\begin{figure}
\epsfxsize=4.5in \centerline{ \epsfbox{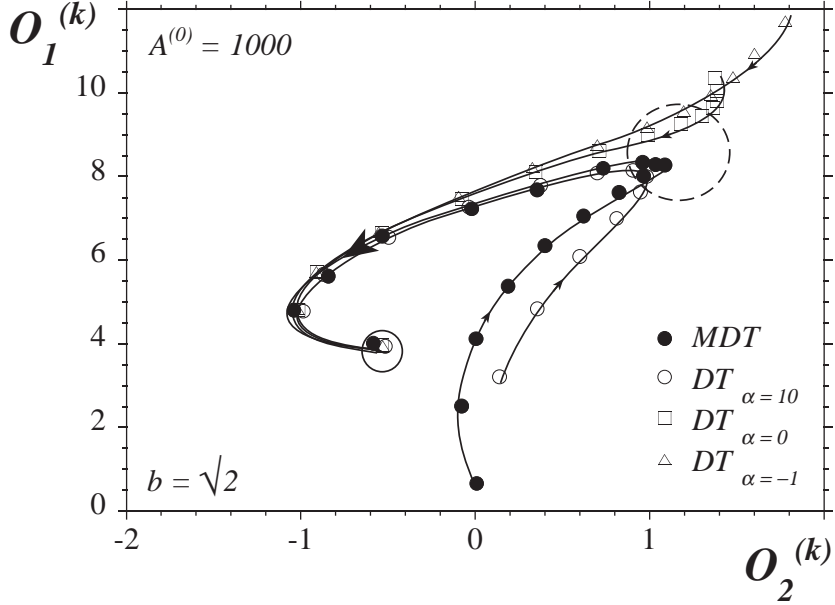}}
\caption{ The RG flow of curvature operators in the $(O_1,
 O_2)${--}plane. The trajectory for minimal triangulations is marked
 with solid circles, while open symbols correspond to combinatorial
 manifolds with varying degrees of coupling to a logarithmic curvature
 term.  The dashed circle indicates a non-trivial fixed point, while
 the solid one is a zero-volume sink.}  \protect\label{fig:op}
\end{figure}

Our findings thus far are consistent with the exact solution of
two-dimensional $R^2$ gravity given in \cite{KSW}.  These authors
solve a matrix model of dually weighted planar graphs which allows the
incorporation of the higher-curvature operator $R^2$. They find that
the $R^2$ operator is non-perturbatively irrelevant in the continuum
limit. The model always reduces at large length scales to a model of
pure gravity {--} there is no transition to a ``flat'' phase. For
technical reasons the solution of \cite{KSW} is restricted to lattices
with {\em even} coordination number vertices. There is thus no direct
mapping to the model we treat. Nevertheless in both cases it is
found that higher-curvature contributions to the effective action 
are irrelevant.

In the second part of this letter we investigate the Ising model
coupled to minimal dynamical triangulations.  In analogy to the
situation in conventional dynamical triangulation we expect that the
addition of Ising spins to the graphs leads to the possibility of a
spin-ordering phase transition at some critical value of the
temperature $\beta_c$ \cite{JKPS}. Performing simulations on triangulations
ranging from 250 to 8000 vertices we indeed observe all the usual
signs of such a transition; a peak in the spin susceptibility which
narrows and grows with increasing lattice volume, a non-zero
magnetization for large coupling and a cusp in the specific heat.

We have attempted to determine the infinite volume critical coupling
$\beta_c$ using an extrapolation of the pseudo-critical coupling
$\beta_c(A)$ associated with the peaks in the specific heat and the
derivative of Binder's cumulant $g_r$. In both cases the location of 
the
peaks is expected to approach $\beta_c$ as $\;|\beta_c - \beta_c(A)|
\sim A^{-1/\nu d_H}$.  This fit is made possible by an independent
determination of $\nu d_H$ {--} the height of the peak in 
$\partial g_r / \partial
\beta$ scales as $A^{1/\nu d_H}$.  In Fig.\ 3 we show the scaling of
the pseudo-critical couplings together with the infinite volume
extrapolation.  Our best estimate of the critical coupling from this
is $\beta_c = 0.2663(3)$.

\begin{figure}
\epsfxsize=4.5in \centerline{ \epsfbox{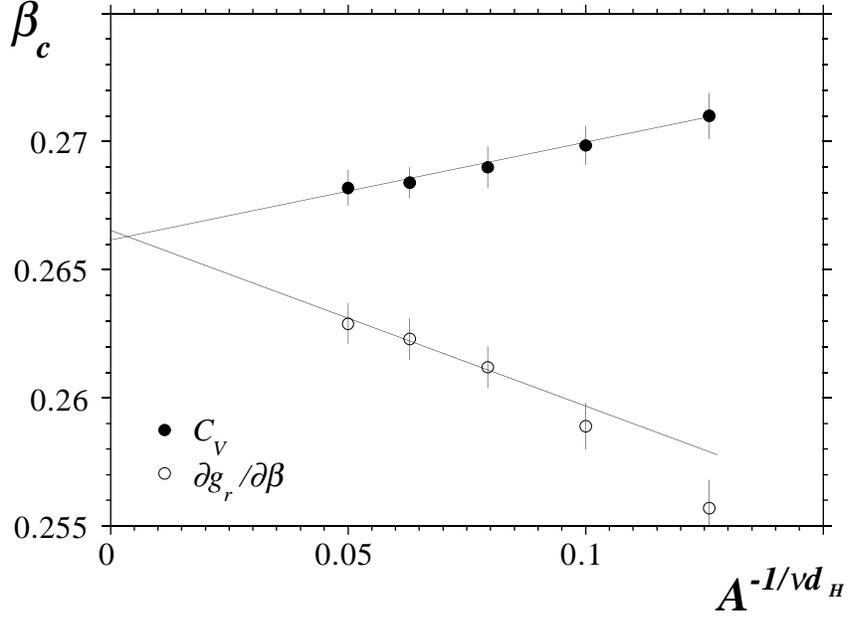}}
\caption{The infinite volume extrapolation of the 
 pseudo-critical couplings obtained from the peaks in
 the specific heat and the derivative of Binder's cumulant.  
 The data points correspond to $A=500, 1000, 2000, 4000$ and $8000$.
 For the Binder's cumulant fit only the three
 largest volumes are included in the extrapolation since finite-size
 effects are more pronounced.
 The rescaling uses the measured value $\nu d_H = 2.9$.}
\protect\label{fig:bc}
\end{figure}

It is interesting to compare this value to the critical 
temperature of the Ising model on other classes of triangulations:
$\beta_c \approx 0.2747$ (fixed triangular lattice),
$\beta_c \approx 0.2163$ (combinatorial),
and $\beta_c \approx 0.1603$ (degenerate). 
It is clear that magnetic ordering occurs more readily
as the triangulation space is enlarged.

To obtain the critical exponents of the model we use
standard finite size scaling.  We have already discussed
$\nu d_H$. In the same way $\alpha$ is obtained from the
scaling of the peak in the specific heat: $C_V =
c_0 + c_1 A^{\alpha/ \nu d_H}$.  From the scaling of
the magnetization ${\cal M}$ and the spin susceptibility
$\chi$ at the critical point we get $\beta$ and $\gamma$
respectively.  The results are shown in Table~\ref{tab2}.  There we
also demonstrate the finite size effects present by systematically
excluding data for small volume.  The results are compared
to the corresponding values for the Onsager solution of the
Ising model on a fixed $2D$ lattice, and the KPZ values for
the Ising model coupled to $2D$-gravity.  It is clear that
for the Ising model coupled to the minimal class of
triangulations the values agree with those of KPZ.
An analytic formulation of a model with vertices of
restricted coordination number has recently been given \cite{wati}.
This model treats squares with coordination number three, four or five.
Although discrete loop equations can be derived they have yet been
solved.

\begin{table}
\begin{center}
\caption{The measured critical exponents. $\nu d_H$ and $\alpha$
 are obtained from the scaling of the peaks in $\partial g_r/
 \partial \beta $ and $C_V$ respectively.  $\beta$ and $\gamma$
 are measured from scaling at $\beta_c$, in which case the errors
 are dominated by the uncertainty in the location of $\beta_c$.}
\vspace{0.1in}
\begin{tabular}{ccccc} \hline
$A$   & $\nu d_H$ & $\alpha$ & $\beta$ & $\gamma$  \\ \hline
250-8000  & 2.857(24) & -0.806(41) & 0.474(20) & 2.117(54) \\
500-8000  & 2.890(33) & -0.922(58) & 0.488(21) & 2.095(52) \\
1000-8000 & 2.907(25) & -0.977(87) & 0.500(18) 
 & 2.070(55) \\ \hline
Onsager   & 2         & 0(log)     & 1/8        & 7/4 \\
KPZ       & 3         & -1         & 1/2        & 2   \\ \hline
\end{tabular}
\label{tab2}
\end{center}
\end{table}

In this letter we have provided strong evidence that a new class of
triangulated models, the MDT models, exhibit the same critical
behavior as the conventional dynamically triangulated random surface
models of two-dimensional quantum gravity. We have demonstrated this
both in pure gravity, where the critical properties are encoded in the
string susceptibility $\gamma_s$, and in the critical Ising model
coupled to gravity where the Ising exponents yield a sensitive test of
the critical behavior of the model. Our results suggest that putting
rather severe cut-offs on the vertex coordination number does not 
affect the universal behavior of the model. This is a highly non-trivial
observation and implies that it may indeed be possible to find
physical $2D$ systems which exhibit this universal
behavior \cite{foot3}.

For quantum gravity it implies that large fluctuations in the
curvature at the scale of the lattice cut-off are unimportant for
determining the continuum structure of $2D$ quantum gravity {--} small
curvature defects are sufficient. The pure gravity model
may be interpreted, similarly to \cite{KSW}, as a gas of strength 
$\pm 1$ curvature defects (or vortices) in flat space.
In the same way the Ising-type model consists of an
interacting gas of Ising spins and curvature defects {--} the
gravitational dressing of the Ising spins being a function of their
interaction with these discrete defects. In this context, it may be
possible to get some insight into the physical origins of this
dressing within this simple model. A study of the geometry of
spin clusters and its relation to the curvature defect distribution
is certainly warranted. 

\vskip  .2in
We would like to thank Konstantinos Anagnostopoulos and David Nelson 
for stimulating discussions. 
This work was supported in part by the Department of
Energy U.S.A under contract No.\ DE-FG02-85ER40237 and from research
funds provided by Syracuse University. Some computations were
performed using facilities at the Northeast Parallel Architectures
Center (NPAC).

\vfill


\begin{thebibliography}{99}
\bibitem{generic}
 F.~David, {\it Simplicial Quantum Gravity and Random Lattices},
  (hep-th/9303127), Lectures given at Les Houches Summer School 
  on Gravitation and Quantization, Session LVII, 
  Les Houches, France, 1992;  
 J.~Ambj\o rn, {\it Quantization of Geometry}, (hep-th/9411179), 
  Lectures given at Les Houches Summer School on Fluctuating 
  Geometries and Statistical Mechanics, Session LXII, 
  Les Houches, France 1994;
 P.~Di Francesco, P.~Ginsparg and J.~Zinn-Justin, Phys.\ Rep.\ 
  254 (1995) 1.
 
\bibitem{frac}
 S.~Catterall, G.~Thorleifsson, M.~Bowick and V.~John, Phys.\
 Lett.\ B354 (1995) 58; J.~Ambj\o rn, J.~Jurkiewicz and 
 Y.~Watabiki, Nucl.\ Phys.\ B454 (1995) 313.

\bibitem{solv}
 V.A.~Kazakov, Phys.\ Lett.\ A119 (1986) 140: 
 D.~Boulatov and V.A.~Kazakov, Phys.\ Lett.\ B186 (1987) 379;
 Z.~Burda and J.~Jurkiewicz, Acta.\ Phys.\ Polon.\ B20 (1989) 949.
 
\bibitem{baby2}
 S.~Jain and S.D.~Mathur, Phys.\ Lett. B286 (1992) 239;
 J.~Ambj\o rn, S.~Jain and G.~Thorleifsson, Phys.\ Lett.\ 
 B297 (1993) 34; J.~Ambj\o rn and G.~Thorleifsson, Phys.\ 
 Lett.\ B323 (1994) 7.  

\bibitem{foot1} Strictly speaking surfaces obtained by cutting along a
 minimal neck are not in ${\cal T}_M$ since they have punctures and
 the vertices on the boundary do not obey the same restrictions as
 those in the interior.  This distinction is not important for our
 numerical measurements.

\bibitem{foot2}
The triangulations are updated using the standard
link-flip algorithm. Its ergodicity on the class of minimal
triangulations is, though, an open issue. The usual proof of ergodicity
for DTRS\cite{BKKM} is based on the reduction of dual $\phi^3$ graphs to a
canonical form with the aid of the appropriate Schwinger-Dyson
equations for amplitudes. This proof does not extend directly to the
minimal class of triangulations. 

\bibitem{BKKM}
 D.V.~Boulatov, V.A.~Kazakov, I.K.~Kostov and A.A.~Migdal,
 Nucl. Phys. B275 (1986) 641. 

\bibitem{ndec}
 G.~Thorleifsson and S.M.~Catterall, Nucl.\ Phys.\ B461 (1996) 350.

\bibitem{KSW}
V.A.~Kazakov, M.~Staudacher and T.~Wynter, 
Nucl.\ Phys.\ B471 (1996) 309.


\bibitem{JKPS}
J.~Jurkiewicz, A.~Krzywicki, B.~Petersson and B.~Soderberg,
Phys. \ Lett. B213 (1988) 511.

\bibitem{foot3}
One such physical system may be a {\em liquid
ferromagnet}. We caution, however, that the Ising model coupled to
minimal dynamical geometry would not be a planar liquid ferromagnet
because of the highly fractal nature of the intrinsic two-dimensional
geometry in this system. It remains a challenge to identify an
appropriate physical system to which our model applies.

\bibitem{wati}S. E. Carroll, M.E. Ortiz and W. Taylor, Nucl.\ Phys.\ B.
468 (1996) 383.

\end{thebibliography}
\end{document}